\documentstyle[seceq,wrapft]{ptptex}

\input epsf

\def\C{{\cal C}}
\def\B{{\cal B}}
\def\R{{\cal R}}
\def\d{{\delta}}
\def\E{{\cal W}}
\def\H{{\cal E}}
\def\A{{\cal A}}
\def\dqE{{\delta q_\E}}

\notypesetlogo
\title{
Brane cosmology: an introduction
}
\author{
David {\sc Langlois}
\footnote{langlois@iap.fr}
}

\inst{
GReCO, Institut d'Astrophysique de Paris
\\
98 bis, Boulevard Arago, 75014 Paris, France}

\recdate{
\today
}

\abst{
These notes give  an introductory  review  on brane cosmology. This subject 
deals with the cosmological behaviour of a brane-universe, i.e. a
three-dimensional space, where ordinary matter is confined, embedded in 
a higher dimensional spacetime. In the tractable case of a five-dimensional 
bulk spacetime, the brane (modified) Friedmann equation is discussed
 in detail, and various other aspects are presented, such as cosmological 
perturbations, bulk scalar fields and systems with several branes.
}

\def\beq{\begin{equation}}
\def\eeq{\end{equation}}

\begin{document}

\maketitle

\section{Introduction}
It has been recently  suggested that there might exist some extra 
spatial dimensions, not in the traditional Kaluza-Klein sense where the 
extra-dimensions are compactified on a small enough radius to evade 
detection in the form of Kaluza-Klein modes, but in a setting where 
the extra dimensions could be large, under the assumption that
 {\it ordinary matter is confined} 
onto a three-dimensional subspace, called {\it brane} (more
precisely `3-brane', referring to the  three spatial dimensions) 
embedded in a larger 
space, called {\it bulk}. 

Altough the idea in itself is not completely new \cite{precursors}, 
the fact that it might be 
connected to recent string theory developments has suscitated a renewed 
interest. In this respect, an inspiring input has been the model 
suggested by Horava and Witten \cite{hw}, sometimes dubbed  
{\it M-theory}, which describes the low 
energy effective theory corresponding to the strong coupling limit 
of $E_8\times E_8$ heterotic string theory. 
This model is associated with an  
eleven-dimensional  bulk spacetime with 11-dimensional supergravity, 
 the eleventh dimension being compactified via a  $Z_2$ orbifold symmetry.
The two fixed points of the orbifold symmetry define two 10-dimensional 
spacetime boundaries, or 9-branes, on which the gauge groups are defined.
Starting from this configuration, one can distinguish three types 
of spatial dimensions: the orbifold dimension, three large dimensions 
corresponding to the ordinary spatial directions and finally six additional 
dimensions, which can be compactified  in the usual Kaluza-Klein way.   
It turns out that the orbifold dimension might be  larger than the 
six Kaluza-Klein extra dimensions, resulting in an intermediary picture 
with a five-dimensional spacetime, two boundary 3-branes, one of which could 
be our universe, and a ``large'' extra-dimension. This model provides 
the motivating framework for many of the brane cosmological models.

This concept of a 3-brane has also been used  in a purely 
phenomenological way by Arkani-Hamed, Dimopoulos and Dvali \cite{add} (ADD)
as a possible solution to the hierarchy problem in particle physics. 
Their setup is extremely simple since they consider a {\it flat} 
$(4+n)$-dimensional spacetime, thus with $n$ compact extra dimensions with, 
for simplicity, a torus topology and a common size $R$. From 
the fundamental (Planck) $(4+n)$-dimensional mass $M_*$, which embodies
the coupling of gravity to matter from the $(4+n)$-dimensional point of view,
one can  deduce the effective four-dimensional Planck mass 
$M_P$, 
either by integrating the Einstein-Hilbert action over the extra dimensions, 
or by using directly Gauss' theorem. One then finds
\beq
M_P^2=M_*^{2+n}R^n.
\eeq
On sizes much larger than $R$, $(4+n)$-dimensional gravity behaves 
effectively as our usual four-dimensional gravity, and the two can be 
distinguished only on scales sufficiently small, of the order of $R$ or 
below.
The simple but crucial remark of ADD is that a fundamental mass $M_*$ of the 
order of the electroweak scale can explain  the huge four-dimensional 
Planck mass $M_P$  we observe, {\it provided the volume in the 
extra dimensions is very large}. Of course, such a large size for the usual 
extra-dimensions \`a la Kaluza-Klein is forbidden by 
collider constraints, but is allowed when ordinary matter is confined to a 
three-brane. In contrast, the constraints on the behaviour of 
gravity are much weaker since the usual Newton's law has been verified 
experimentally only down to a fraction of millimeter\cite{grav_exp}. 
This leaves room for extra dimensions as large as this millimeter 
experimental bound.

The treatment of extra-dimensions can be refined by allowing the  
spacetime to be curved, by the presence of the brane  and possibly 
by the bulk. In this spirit, Randall and Sundrum have proposed a 
very interesting
 model\cite{rs99a,rs99b} with an Anti de Sitter (AdS) five-dimensional 
spacetime (i.e. a maximally symmetric spacetime with a negative 
cosmological constant), and shown that, for an appropriate tension of 
the brane representing our universe, one recovers effectively 
four-dimensional gravity even with an {\it infinite extra-dimension}. 
Another model \cite{dgp}, which will not be discussed in this review, is 
characterized, in contrast with the previous ones, 
 by a gravity which becomes five-dimensional at {\it large} scales and it
could have interesting applications for the present 
cosmological acceleration\cite{deffayet}.

The recent concept of extra-dimensions with branes has been explored
in  a lot of aspects of particle physics, gravity, astrophysics and cosmology
\cite{maartens,rubakov}.
The purpose of this review is  to present  a very specific, 
although very active,  facet of this vast domain, dealing
with the cosmological behaviour of brane models when the curvature 
of spacetime along the extra-dimensions, and 
in particular the brane self-gravity, is explicitly taken into account. 
For technical reasons,  this can be easily studied in the context 
of codimension one spacetimes and we will therefore restrict this review
to the case of {\it five-dimensional} spacetimes.
Even in this restricted area, the number of works during the last few years is
so large that this introductory review is not intended to be comprehensive 
but will focus on some selected aspects. This also implies that the 
bibliography is far from exhaustive.

\section{Pre-history of brane cosmology} 
 Since our purpose is to describe (homogeneous) cosmology within the brane,
we will model our four-dimensional universe as an infinitesimally 
thin wall of constant spatial curvature embedded in a 
five-dimensional spacetime.
This means that we need to consider spacetimes 
with planar (or spherical/hyperboloidal) symmetry along three 
of their spatial directions, i.e. 
homogeneous and isotropic along three  spatial dimensions. 
The general metric compatible with these symmetries can be written in 
appropriate coordinate systems in the form
\beq
ds^2= g_{ab}(z^c)dz^a dz^b+ a^2(z^c)\gamma_{ij}dx^idx^j,
\label{gen_metric}
\eeq
where $\gamma_{ij}$ is the maximally symmetric three-dimensional metric, 
with either negative, vanishing or positive  spatial curvature (respectively
labelled by 
$k=-1$, $0$ or $1$), and 
$g_{ab}$ is a two-dimensional metric (with space-time signature) 
depending only on the two coordinates $z^c$, which span time and the 
extra spatial dimension. 
We are  interested in the history of a spatially homogeneous 
and isotropic three-brane, which can be simply described as a point
trajectory in the two-dimensional $z$-spacetime, the ordinary 
spatial dimensions inside the brane corresponding to the coordinates 
$x^i$. Now, once such a trajectory has been defined, it is always 
possible to introduce a so-called Gaussian Normal (GN) coordinate 
system so that 
\beq
g_{ab}dz^a dz^b=-n(t,y)^2 dt^2+dy^2.
\eeq
This coordinate system can be constructed by introducing the proper time 
on the trajectory and by defining the coordinate $y$ as the proper distance 
on (space-like) geodesics normal to the trajectory. 
The time coordinate defined only on the brane is then propagated off 
the brane along these normal geodesics.
It is convenient 
to take the origin on  the brane so that $y=0$ in this coordinate system
represents   the brane itself. 

Summarizing, it is always possible to write the metric (\ref{gen_metric}), 
in a 
GN coordinate system where the brane is located at $y=0$, in the form
\beq
ds^2=- n(t,y)^2 dt^2+a(t,y)^2 \gamma_{ij}dx^idx^j+dy^2.
\label{metric}
\eeq
This is the most convenient system of coordinates when one wishes to 
focus on the brane itself since the induced metric in the brane is immediately 
obtained in its Friedmann-Lema\^\i tre-Robertson-Walker (FLRW) form 
\beq
ds^2_{brane}=-n(t,0)^2 dt^2+a(t,0)^2 \gamma_{ij}dx^idx^j.
\eeq
If $t$ is the proper time on the brane then $n(t,0)=1$.
The GN coordinates can however suffer from coordinate singularities off 
the brane, and it might be more convenient, as discussed in Section 5,
to use more appropriate coordinates in order to describe the bulk 
spacetime structure.

Having specified the form of the metric, we now turn to the five-dimensional 
Einstein equations, which can be written in the condensed form
 \beq
G_{AB}\equiv R_{AB}-{1\over 2}Rg_{AB}= -\Lambda g_{AB}+\kappa^2 T_{AB},
\label{einstein}
\eeq
where $R_{AB}$ is the five-dimensional Ricci tensor and $R=R_A^A$ its trace.
$\Lambda$ stands for a possible bulk cosmological constant, whereas 
$T_{AB}$ is the total energy momentum tensor. It includes the energy-momentum 
tensor of the brane, which is distributional if one assumes the brane 
infinitely thin (a few works  \cite{kkop99,ml02} have discussed the 
cosmology for  thick branes, which can also be related to the abundant 
literature on thick domain walls\cite{cvetic}) 
and thus of the form
\begin{equation}
T^A_{\quad B}|_{_{\rm brane}}=S^A_{\quad B} \delta (y)= \mbox{diag} 
\left(-\rho_b,p_b,p_b,p_b,0 \right)\delta (y),
\label{source}
\end{equation}
where $\rho_b$ is the total brane energy density and $p_b$ is the total 
brane pressure. The total energy-momentum tensor also includes a possible 
bulk contribution, which we first assume  to vanish (bulk matter will be 
considered in Section 8).

In the coordinate system (\ref{metric}), the components of the Einstein 
tensor read\cite{bdl99}
\begin{eqnarray}
{ G}_{00} &=& 3\frac{\dot{a}^2}{a}  - 3{n^2} 
\left({a''\over a}+{{a'}^2\over a^2}\right) 
 +3 k \frac{n^2}{a^2}, 
\label{ein00} \\
 { G}_{ij} &=& 
a^2 \gamma_{ij}\left(2\frac{a^{\prime \prime}}{a}
+\frac{n^{\prime \prime}}{n}+ {{a'}^2\over a^2}
+2{a^\prime n^\prime\over an}\right)
\nonumber \\
& &+\frac{a^2}{n^2} \gamma_{ij} \left(
-2{\ddot a\over a}- {\dot{a}^2\over a^2}+2{\dot{a}\dot{n}\over an}
\right)
-k \gamma_{ij},
\label{einij} \\
{ G}_{0y} &=&  3\left({n^\prime\over n} {\frac{\dot{a}}{a}}  - 
\frac{\dot{a}^{\prime}}{a}
 \right),
\label{ein05} \\
{G}_{yy} &=& 3\left( {{a'}^2\over a^2}
+{a^\prime n^\prime\over an}  \right) 
- \frac{3}{n^2} 
\left({\ddot a\over a}+{\frac{\dot{a}^2}{a^2}} -\frac{\dot{a}}{a}
\frac{\dot{n}}{n} \right) - 3 \frac{k}{a^2},
\label{ein55} 
\end{eqnarray} 
where a dot stands for a derivative with respect to $t$ and 
a prime  a derivative with respect to $y$.
One way to solve Einstein's equations is to insert the full energy-momentum
tensor, including the brane energy-momentum tensor with its 
delta distribution,  on the right-hand side of (\ref{einstein}) 
and solve  the full system 
of equations. An alternative procedure 
is to consider  Einstein's  equations without the brane energy-momentum 
tensor, i.e. valid strictly in the bulk, and then  impose 
on the general solutions boundary conditions to take into account the 
physical presence of the brane. The Einstein tensor is made of the metric 
up to second derivatives, so formally the Einstein equations with the 
distributional source are of the form
\beq
g''=S\ \delta(y).
\eeq
If the brane is located at $y=0$, integrating this equation over $y$ across
the brane immediately yields
\beq
g'(+\epsilon)-g'(-\epsilon)=S.
\label{formal_jump}
\eeq
In other words, the boundary conditions due to 
 the presence of the brane must take the form of a relation
between the jump, across the brane,  of the first derivative of the metric 
with respect to $y$ and the brane energy-momentum tensor.
In the case of Einstein's equations, the exact equivalent of the formal
 relation (\ref{formal_jump}) 
is the so-called junction conditions \cite{israel}, 
which can be written in a 
covariant form  as
\beq
 \left[K^A_{\, B}
-K\delta ^A_{\, B}\right]=-\kappa^2 S^A_{\, B},
\label{israel}
\eeq
where the brackets here denote the jump at the 
brane, i.e. $[Q]=Q_{\{y=0^+\}}-Q_{\{y=0^-\}}$, and the extrinsic curvature 
tensor is defined by 
\beq
K_{AB}=h_{A}^C\nabla_C n_B,
\label{extrinsic}
\eeq
$n^A$ being the unit vector normal to the brane, and 
\beq
h_{AB}=g_{AB}-n_An_B
\eeq
the induced metric on the brane.

A frequent assumption in the brane cosmology literature has been, for 
simplicity,  to keep  the orbifold nature of the extra dimension 
in the Horava-Witten model 
and thus impose a mirror symmetry across the brane, although 
some works have relaxed this assumption \cite{stw00,bcmu01}. 
This enables us to 
replace the jump in the extrinsic curvature by twice the value of the 
extrinsic curvature at the location of the brane.
The junction conditions 
(\ref{israel}) then imply
\beq
K_{AB}=-{\kappa^2\over 2} \left(S_{AB}-{1\over 3}S g_{AB}\right),
\label{junction}
\eeq
where $S\equiv g^{AB} S_{AB}$ is the trace of $S_{AB}$. 
If one uses the GN coordinate system with the metric (\ref{metric}) and 
substitutes the explicit form of the brane energy-momentum tensor
 (\ref{source}), the 
junction conditions reduce to the two relations 
\beq
\left({n'\over n}\right)_{0^+}={\kappa^2\over 6}\left(3p_b+2\rho_b\right),
\qquad
\left({a'\over a}\right)_{0^+}=-{\kappa^2\over 6}\rho_b.
\label{junction_a_n}
\eeq
Going back to the bulk Einstein equations, one can solve the  
component $G_{0y}=0$ (see \ref{ein05}) to get
\beq
\dot a(t,y)=\nu(t)\, n(t,y), 
\eeq
and the integration of the component $(0-0)$ with respect to $y$ and 
of the component $(y-y)$ with respect to time  yields the  first integral
\beq
(aa')^2-\nu^2 a^2 -ka^2+{\Lambda\over 6} a^4+\C=0,
\eeq
where $\C$ is an arbitrary integration constant. 
When one evaluates this first integral at $y=0$, i.e. in our brane-universe, 
substituting the junction conditions given above in (\ref{junction}), 
one ends up with the following equation\cite{bdl99,bdel99}
\beq
H^2\equiv {\dot a_0^2\over a_0^2}={\kappa^4\over 36}\rho_b^2+{\Lambda\over 6}
-{k\over a^2}+{\C\over a^4}.
\label{fried}
\eeq
where the subscript `$0$' means evaluation at $y=0$.
This equation relates the Hubble parameter to the 
energy density  but it is different from the usual Friedmann equation
[$H^2=(8\pi G/3)\rho$]. 
The most  remarkable feature of (\ref{fried}) 
 is that {\it the energy density of the brane enters 
quadratically} on the right hand side in contrast with the standard 
four-dimensional Friedmann equation where the energy density enters 
linearly. 
As for  the  energy conservation equation,  it 
is unchanged in this five-dimensional setup
and still reads
\beq
\dot\rho_b+3H(\rho_b+p_b)=0,
\label{conservation}
\eeq
as a consequence of the component $(0-y)$ of Einstein's equations 
combined with the junction conditions (\ref{junction_a_n}).

In the simplest case where $\Lambda=0$ and $\C=0$, 
one can easily solve the  cosmological equations  
(\ref{fried}-\ref{conservation}) 
for a perfect fluid with  equation of state $p_b=w\rho_b$ ($w$ constant).
One finds that the evolution of the scale factor is given by\cite{bdl99}
\beq
a_0(t)\propto t^{1\over 3(1+w)}.
\label{a_unconventional}
\eeq
In the most interesting cases for cosmology, radiation and pressureless 
matter, the evolution of the scale factor is thus given  by, respectively,  
$a\sim t^{1/4}$  (instead 
of the usual $a\sim t^{1/2}$) and $a\sim t^{1/3}$  (instead 
of  $a\sim t^{2/3}$).
Such behaviour  is problematic because it cannot be reconciled 
with nucleosynthesis. Indeed, the standard 
nucleosynthesis scenario depends crucially 
on the  balance between  the microphysical  reaction rates
 and the expansion rate of the universe. Any drastic change in the  
 evolution of the scale factor between nucleosynthesis and now 
 would dramatically modify 
the predictions for the light element abundances.  
After discussing some  particular solutions in the next section, 
the subsequent section will  present a brane 
cosmological model with much nicer features.

\section{Cosmological solutions for `domain walls'}
It is instructive to consider
 the brane cosmological solutions for the simplest 
equation of state, $p_b=-\rho_b$, which also characterizes the so-called 
  domain walls.    
With this particular equation of state, 
the cosmological equations can be explicitly integrated \cite{kaloper99,nihei99,bcg00}.
Indeed, the energy density is necessarily 
constant, as implied by (\ref{conservation}), and 
 the Friedmann equation (\ref{fried}) is  of the form
\beq
{\dot a^2\over a^2}=\alpha-{k\over a^2}+{\C\over a^4},
\label{fried_dw}
\eeq
where
\beq
\alpha\equiv  {\kappa^4\over 36}\rho_b^2+{\Lambda\over 6}
\eeq
is a constant. The case $\alpha=0$ is referred to as a  `critical' brane 
(or domain wall),
while $\alpha<0$ and $\alpha>0$ correspond to  subcritical and supercritical 
branes respectively.
Defining $X\equiv a^2$, the Friedmann equation 
(\ref{fried_dw}) can be rewritten as
\beq
{\dot X^2\over 4}=\alpha X^2-kX+\C,
\label{eqX}
\eeq
which is analogous to the equation for a point particle with kinetic energy
on the left hand side and (minus) the potential energy on the right hand side.
For a critical brane ($\alpha=0$), one immediately 
finds the following three solutions, depending on the spatial 
curvature of the brane:
\begin{eqnarray}
a&=\sqrt{2}\ {\C}^{1/4} \, t^{1\over 2},  &\qquad (k=0), \\
a&=\left(t^2+2\sqrt{\C}\, t\right)^{1/2}, &\qquad (k=-1), \\
a&=\sqrt{\C-t^2}, &\qquad (k=+1),
\end{eqnarray}
which in fact correspond to the usual FLRW solutions with radiation, 
the term proportional to $\C$ playing the effective r\^ole of radiation.

For a non critical brane, 
integration of  (\ref{eqX}) yields the following solutions, 
\begin{eqnarray}
a^2&=\sqrt{\beta\over\alpha}\sinh(2\sqrt{\alpha}t)+{k\over 2\alpha} 
&\quad (\alpha>0, \quad \beta>0) 
\label{++}\\
a^2&=\pm \sqrt{-\beta\over\alpha}\cosh(2\sqrt{\alpha}t)+{k\over 2\alpha} 
&\quad (\alpha>0, \quad \beta<0) 
\label{+-}\\
a^2&=\sqrt{-\beta\over\alpha}\sin(2\sqrt{\alpha}t)+{k\over 2\alpha} 
&\quad (\alpha<0, \quad \beta>0),
\label{-+}
\end{eqnarray}
with 
\beq
\beta=\C-{k^2\over 4\alpha}.
\eeq
The particular solutions for $\beta=0$ (and $\alpha>0$)
are
\beq
a^2={k\over 2\alpha}\pm e^{\pm 2\sqrt{\alpha} t}.
\label{+0}
\eeq 
In all the above equations the (additive) 
integration constant defining the origin of times is not written explicitly 
and one can always use this time shift to rewrite any solution in 
a more adequate form (for instance so that $a=0$ at $t=0$). Moreover, all 
solutions have their time reversed counterpart obtained by changing $t$ into 
$-t$. 

The  analytical expressions (\ref{+-}) or (\ref{-+}) cover very 
different cosmological behaviours depending on the value of their 
parameters. Using  
 the analogy with the point particle  mentioned
before, one can see that the global cosmological behaviour depends 
on the number of positive roots of the quadratic `potential' on  the 
right hand side of (\ref{eqX}).  
For $\alpha>0$, $k=1$ and  $0<\C<1/(4\alpha)$, the potential has  two positive 
roots and this leads to two solutions, one  expanding 
from the singularity 
$a=0$ and recontracting to $a=0$, the other  contracting from 
infinity to a minimum scale factor and then expanding. These two types 
of solutions correspond to the two branches of (\ref{+-}).
For $\alpha>0$ and $\C>0$ with $k=0,-1$, or $\C>1/(4\alpha)$ with $k=1$, 
there is no positive root and the corresponding cosmology starts from 
the singularity $a=0$ and then expands indefinitely. This behaviour 
is described by the solution (\ref{++}) and the solution 
(\ref{+-}) for $k=-1$. 
For $\alpha<0$ and $\C>0$, there is a single positive root, and the 
cosmological solution starts expanding from $a=0$ and then recollapses back 
to $a=0$. This corresponds to the solution (\ref{-+}).
For $\alpha<0$, $k=-1$ and $1/(4\alpha)<\C<0$, the solution is confined 
between the two positive roots, which gives an oscillating cosmology between 
a minimum scale factor and a maximum scale factor. This is also described 
by the solution (\ref{-+}).
Finally, for $\alpha >0$ and $\C<0$, there is a single positive root and 
the solution corresponds to a contraction from infinity followed, after 
a bounce at a minimum scale factor, by an expansion back to infinity. This 
is described by the solution (\ref{+-}).

\section{Simplest realistic brane cosmology}
As explained earlier, the simplest model of self-gravitating 
brane cosmology, that of a brane embedded in an empty  bulk 
with vanishing cosmological constant (in fact a Minkowski bulk 
because of the symmetries), does not appear compatible \cite{bdl99} 
with  the standard landmarks of modern cosmology. It is thus necessary 
to consider 
 more sophisticated models in order to get a viable scenario, at 
least as far as homogeneous cosmology is concerned.

An instructive exercise is to look for non trivial (i.e. with 
a non empty brane) {\it static} solutions. In the simplest case $\C=0$, one 
immediately sees from (\ref{fried}) 
that  a static solution, corresponding to $H=0$, can be 
obtained with a {\it negative cosmological constant} and an 
energy density $\rho_b$ satisfying
\beq
\kappa^2\rho_b=\pm\sqrt{-6\Lambda}.
\label{rs}
\eeq
One can check that this is compatible with the other equations, in particular
the conservation equation in the brane (\ref{conservation}), 
which imposes that the matter 
equation of state is that of a cosmological constant, i.e. $p_b=-\rho_b$.
It turns out that this configuration is  exactly the starting point  of
 the two models due to Randall and Sundrum (RS) \cite{rs99a,rs99b}. 
From the point of view 
of brane cosmology, embodied in the unconventional Friedmann equation 
(\ref{fried}), the RS models thus 
appear as the simplest non trivial {\it static}
brane configurations. The case of the single brane model\cite{rs99b}, with 
a positive tension, is particularly interesting, because ordinary 
four-dimensional gravity, at least at linear order,
 is effectively recovered on large enough lengthscales \cite{gt99}, 
with 
\beq
 8\pi G\equiv M_P^{-2}=\kappa^2\ell^{-1},
\label{G_eff}
\eeq
where $\ell$ is the Anti de Sitter (AdS) lengthscale defined by the (negative) cosmological constant, 
\beq
\Lambda=-{6\over \ell^2}.
\eeq
Since usual gravity is recovered,  the generalization of this model 
to cosmology   seems a priori a good candidate 
for a viable brane cosmology.

Let us  therefore consider a  brane with the 
 total energy density 
\beq
\rho_b=\sigma+\rho,
\label{sum}
\eeq
where $\sigma$ is a tension, constant in time, and $\rho$
the  energy
density of ordinary cosmological matter. 
Substituting this decomposition into (\ref{fried}), one obtains
\cite{cosmors,bdel99,ftw99}
\beq
H^2= \left({\kappa^4\over 36}\sigma^2-\mu^2\right)
+{\kappa^4\over 18}\sigma\rho
+{\kappa^4\over 36}\rho^2-{k\over a^2}+{\C\over a^4},
\label{bdel}
\eeq
where $\mu\equiv\ell^{-1}$ is the AdS mass scale. 
If one fine-tunes the brane tension and the bulk cosmological cosmological 
constant like in (\ref{rs}) so that 
\beq
{\kappa^2\over 6}\sigma=\mu,
\label{rs_bis}
\eeq
the first term on the right hand side of (\ref{bdel}) vanishes and, because 
of (\ref{G_eff}), the tension is proportional to Newton's constant,
\beq
\kappa_4^2\equiv 8\pi G= {\kappa^4\over 6}\sigma={\kappa^2}\mu.
\label{newton}
\eeq
The second term in (\ref{bdel}) 
then becomes the dominant  term if $\rho$ is small enough and
is {\it exactly the linear term of the usual Friedmann equation}, with the 
same coefficient of proportionality.

The third term on the right hand side of (\ref{bdel}), 
{\it quadratic in the energy density}, 
provides a {\it high-energy correction} to the Friedmann equation 
which becomes significant when the value of the energy density approaches 
the value of the tension $\sigma$ and dominates  at still higher 
energy densities. In the very high energy regime, $\rho\gg \sigma$, one 
 recovers the unconventional behaviour (\ref{a_unconventional}), 
not surprisingly since the 
bulk cosmological constant is then  negligible.

Finally, the last term in (\ref{bdel}) behaves like radiation and 
arises from the integration constant $\C$. This 
constant $\C$ is  analogous to the Schwarzschild mass, as will be shown in the 
next section,  and it is  
related to the bulk Weyl tensor, which vanishes when $\C=0$. In a 
cosmological context, this term is constrained to be small enough 
at the time of nucleosynthesis in order to satisfy the constraints on the 
number of extra light degrees of freedom (this will be discussed 
quantitatively just below). In the matter era, this term 
then redshifts quickly and would be in principle negligible today.

To summarize the above results,  
the brane Friedmann equation (\ref{fried}), for a RS type 
brane (i.e. satisfying (\ref{sum}) and (\ref{rs_bis})),  reduces to
\beq
H^2={8\pi G\over 3}\rho\left(1+{\rho\over 2\sigma}\right)+{\C\over a^4},
\label{fried_bis}
\eeq
which shows that, {\it at low energies, i.e. at late times, one recovers 
the usual Friedmann equation}. Going backwards in time, the $\rho^2$ 
term becomes significant and makes brane cosmology deviate from the usual 
FLRW behaviour.

It is also useful, especially as a preparation to the equations for 
the cosmological perturbations, (see Section 7), to notice that 
the generalized Friedmann 
equation  (\ref{fried_bis}) can be seen as a particular case of 
the  more general 
effective four-dimensional Einstein's equations for the brane metric 
$g_{\mu\nu}$,  obtained by 
projection on the brane. Using the Gauss equation and the junction
conditions, and decomposing 
the total energy-momentum tensor of the brane into a pure tension part and 
an ordinary  matter part so that 
\beq
S_{\mu\nu}=-\sigma g_{\mu\nu}+\tau_{\mu\nu},
\eeq
one arrives to  effective four-dimensional Einstein equations, which read 
\cite{sms99}
\beq
G_{\mu\nu}+\Lambda_4 g_{\mu\nu}=\kappa_4^2\tau_{\mu\nu}+\kappa^2\Pi_{\mu\nu}
-E_{\mu\nu},
\label{einstein4d}
\eeq
with 
\beq
\Lambda_4 ={\Lambda\over 2}+{\kappa^4\over12}\,\sigma^2  \,, \qquad
\Pi_{\mu\nu}=
-\frac{1}{4} \tau_{\mu\alpha}\tau_\nu{}^{\alpha}
+\frac{1}{12}\tau \tau_{\mu\nu}
+\frac{1}{24}\left(3\tau_{\alpha\beta}\tau^{\alpha\beta}-\tau^2\right)
g_{\mu\nu}\,,
\label{pi}
\eeq
and where 
\beq
E_{\mu\nu}=C^y_{\ \mu y \nu} 
\eeq
is the projection on the brane of the five-dimensional Weyl
tensor.
In this new form, the effective Einstein's equations are analogous 
the standard four-dimensional
equations with the replacement of the usual matter energy-momentum tensor 
$\tau_{\mu\nu}$ by the sum of an {\it effective matter energy-momentum tensor},
\beq
T_{\mu\nu}^{eff}=\tau_{\mu\nu}+{\kappa^4\over \kappa_4^2}\Pi_{\mu\nu},
\eeq 
which is constructed  only with $\tau_{\mu\nu}$, and of an additional 
part that depends on the bulk, which defines what we will call the 
{\it Weyl energy-momentum tensor}
\beq
T^{Weyl}_{\mu\nu}=-\kappa_4^{-2}E_{\mu\nu}.
\label{Tweyl}
\eeq
Although this formulation
might appear very simple, the reader should be warned  that this equation is in
fact  directly useful only in the cosmological case, where $E_{\mu\nu}$
reduces to the arbitrary constant $\C$. In general, $E_{\mu\nu}$ will hide
a dependence on   the brane content and only a detailed study of the bulk 
can in practice provide the true behaviour of gravity in the brane.

Let us now work out a few explicit cosmological solutions. For an equation 
of state $p=w\rho$, with $w$ constant, one can integrate explicitly the 
conservation equation (\ref{conservation}) to obtain as usual
\beq
\rho=\rho_0 a^{-q}, \quad q\equiv 3(w+1).
\eeq
Substituting in the Friedmann equation (\ref{bdel}), one gets
\beq
{\dot a^2\over a^2}= \alpha
+{\kappa^4\over 18}\sigma\rho_0 a^{-q}
+{\kappa^4\over 36}\rho_0^2 a^{-2q} +{\C\over a^4}-{k\over a^2}.
\label{fried_q}
\eeq
with 
\beq
\alpha= {\kappa^4\over 36}\sigma^2-\mu^2.
\eeq
In the case $\C=0$ and $k=0$, defining $X\equiv a^q$, the Friedmann equation
(\ref{fried_q}) reduces to the form
\beq
{\dot X^2\over q^2}=\alpha X^2+\beta X+\xi,
\qquad \beta={\kappa^4\over 18}\sigma\rho_0, \quad
\xi= {\kappa^4\over 36}\rho_0^2,
\eeq
which is similar to  (\ref{eqX}). 
For a critical brane, i.e. $\alpha=0$, the corresponding cosmological 
evolution is given by
\beq
a^q={q^2\over 4}\beta t^2+ q\sqrt{\xi}t.
\label{a_brane}
\eeq
It is clear from this analytical expression  that 
there is a transition, at a typical time of the order of 
$\mu^{-1}$, i.e. of the  AdS lengthscale $\ell$, 
 between an early high energy regime characterized by the behaviour 
$a\propto t^{1/q}$ and 
a late low energy regime characterized by the standard evolution 
$a\propto t^{2/q}$.

For a non critical brane, the cosmological solutions are 
similar to the expressions (\ref{+-}) and (\ref{-+}) [the expression
(\ref{++}) does not apply here because of the relation 
between the coefficients $\alpha$, $\beta$ and $\xi$].
After choosing an appropriate origin of time (so that $a(0)=0$), they 
can be rewritten in the form
\beq
a^q=\sqrt{\xi\over\alpha}\sinh\left(q\sqrt{\alpha}t\right)
+{\beta\over 2\alpha}\left[\cosh\left(q\sqrt{\alpha}t\right)-1\right]
\qquad (\alpha>0)
\label{a_dS}
\eeq
 and 
\beq
a^q=\sqrt{\xi\over -\alpha}\sin\left(q\sqrt{\alpha}t\right)
+{\beta\over 2\alpha}\left[\cos\left(q\sqrt{\alpha}t\right)-1\right]
\qquad (\alpha <0).
\eeq
The solution (\ref{a_dS}) describes a brane cosmology with 
a positive cosmological constant. This implies that, by fine-tuning 
adequately the brane tension, one can obtain a cosmology with, like 
in (\ref{a_brane}),  
an unconventional early phase  followed by a conventional phase, itself
followed by a period dominated by a cosmological constant.
\begin{figure}
\epsfbox{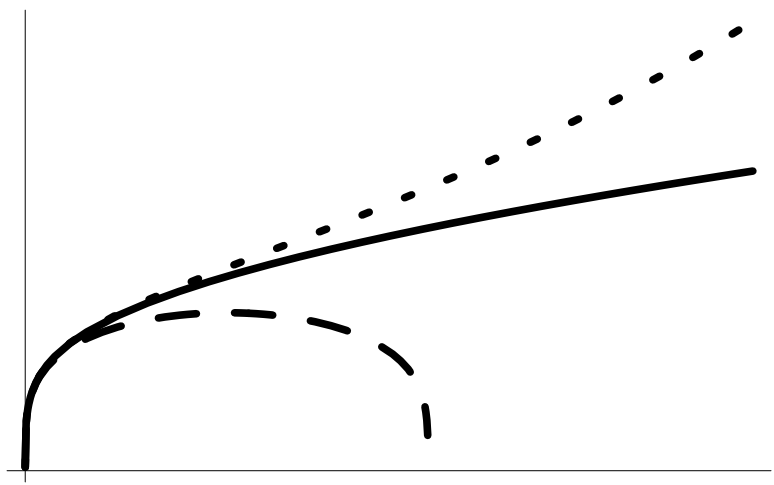} 
\caption{Evolution of the scale factor for $\alpha=0$ (continuous line), 
$\alpha>0$ (dotted line) and $\alpha<0$ (dashed line)}
\end{figure}

In the present section,  we have so far considered only  the metric 
in the brane. But the metric  outside the brane can be also 
determined explicitly by solving the full system of Einstein's equations.
\cite{bdel99}.
The metric coefficients defined in (\ref{metric})  are given by 
\beq
a(t,y)={a_0(t)\over\sqrt{2}}\left\{(1-\eta^2)-{\C\over \mu^2 a_0^4}+
\left[(1+\eta^2)+{\C\over \mu^2 a_0^4}\right]\cosh(2\mu y)
-2\eta \sinh(2\mu |y|)\right\}^{1/2}
\label{a_bulk}
\eeq
and 
\beq
n(t,y)={\dot a(t,y)\over \dot a_0(t)},
\label{n_bulk}
\eeq
which, in the special  case $\C=0$, reduce to the much simpler expressions 
\begin{eqnarray}
a(t,y)&=& a_0(t)\left(\cosh\mu y-\eta \sinh\mu|y|\right)\\
n(t,y)&=& \cosh\mu y-\tilde\eta \sinh\mu|y|,
\label{bulk_metric}
\end{eqnarray}
with
\beq
\eta=1+{\rho\over\sigma}, \qquad \tilde\eta=\eta+{\dot\eta\over H_0}.
\label{eta}
\eeq
In the  limit $\rho=0$, i.e. $\rho_b=\sigma$, which 
implies  $\eta=\tilde\eta=1$, one recovers the RS metric
$a(y)=n(y)=\exp(-\mu|y|)$.

In summary, we have obtained a cosmological model, based on a braneworld
scenario, which appears to be 
compatible with current observations, because it converges to the 
standard model  at low enough energies. 
Let us now quantify the constraints on the parameters 
of this model. 
As mentioned above an essential  constraint  comes 
from nucleosynthesis: the evolution of the universe since
nucleosynthesis must be  approximately 
the same as in usual cosmology. This is the case if 
the energy scale associated with the tension is higher than 
the nucleosynthesis
energy scale, i.e.
\beq
M_c\equiv\sigma^{1/4} > 1 \ {\rm MeV}.
\eeq
Combining this with (\ref{newton}) this implies for the fundamental mass scale
(defined by $\kappa^2\equiv M^{-3}$)  
\beq
M > 10^4 \ {\rm GeV}.
\eeq
However, one must not forget another constraint, not of cosmological nature:
the requirement to recover ordinary gravity down to scales of the 
submillimeter order. This requires\cite{grav_exp}
\beq
\ell <  10^{-1} \ {\rm mm},
\eeq
which yields in turn the constraint
\beq
 M >  10^8 \ {\rm GeV}.
\eeq
Therefore the most stringent constraint comes, not from cosmology, but
from gravity experiments in this particular model.

There are also constraints on the Weyl parameter $\C$\cite{bdel99,others}. 
The ratio
\beq
\epsilon_W={\rho_{Weyl}\over \rho_{rad}}={\C\sigma\over 2\mu^2\rho_{rad}a^4}
\eeq
is constrained by the number of additional relativistic degrees of freedom 
allowed during nucleosynthesis \cite{osw99}, which is usually expressed 
as the number of additional light neutrino species  $\Delta N_\nu$. 
A typical bound $\Delta N_\nu<1$ implies $\epsilon_W<8\times 10^{-2}$ at 
the time of nucleosynthesis. It is also important to stress that if 
$\C$ has been considered here as an arbitrary {\it constant}, a more 
refined, and more realistic, treatment must take into account the fact that 
cosmology is only approximately homogeneous: the small inhomogeneities in the
brane can produce bulk gravitons and the energy outflow carried by these
gravitational waves can feed the asymptotic `Schwarzschild mass', i.e. the 
Weyl parameter $\C$ which would not be any longer a constant 
(in agreement with the fact that the bulk is then 
no longer vacuum). This process
is at present under investigation and the first estimates give a 
Weyl parameters of the order of the nucleosynthesis limit 
\cite{hm01,lsr02}.

So far, we have thus been able to build a model, which reproduces all 
qualitative and quantitative features of ordinary cosmology in the 
domains that have been tested by observations. 
The obvious next question is whether this will still hold for a more 
realistic cosmology that includes  perturbations  from homogeneity, and more 
interestingly, whether  
brane cosmology is capable of 
providing  predictions that  deviate from usual cosmology and 
which might tested in the  future. This question will be addressed, but 
unfortunately not answered, in the section following  the next one which 
presents the   brane cosmological solutions in a totally different way.

\section{A different point of view}
In the previous sections, we have from the start restricted ourselves to 
a particular system of coordinates, namely a GN coordinate system. 
This choice entails  no  physical restriction (at least locally) and 
is very useful for a `brane-based' point of view, but there also 
exists an alternative approach for deriving the brane cosmological 
solutions given earlier, which relies on a `bulk-based' point of 
view \cite{kraus99,ida99} and corresponds to a more appropriate choice of 
coordinates for the bulk. 
The link between the two points of view can be understood from 
a general analysis \cite{bcg00} which consists  in solving 
the five-dimensional  Einstein equations for a generic metric 
of the form (\ref{gen_metric}). 
It turns out that simple coordinates  emerge, 
expressing in a manifest way the underlying symmetries 
of the solutions. This  approach leads to the following simple form for 
the solutions (with the required `cosmological 
symmetries')  of Einstein's  equations in the bulk (\ref{einstein}) [with 
$T_{AB}=0$]:
\beq
ds^2=-f(R)\, dT^2+{dR^2\over f(R)}+R^2\, \gamma_{ij}\, dx^i dx^j, 
\label{adsmetric}
\eeq
where 
\beq
f(R)\equiv k-{\Lambda\over 6} R^2-{\C\over R^2}.
\label{f}
\eeq
The above metric is known as the five-dimensional Schwarzschild-Anti de 
Sitter (Sch-AdS) metric 
(AdS for $\Lambda<0$, which is the case we are interested in; for 
$\Lambda>0$, this the Schwarschild-de Sitter metric). 
It is clear from (\ref{f}) that $\C$ is indeed the five-dimensional 
analog of the Schwarschild mass, as said before (the $R^{-2}$ dependence
 instead of the usual $R^{-1}$ is simply due to the different
 dimension of spacetime).

It is manifestly static (since the metric 
coefficients are time-independent), which means that the solutions 
of  Einstein's equations have more symmetries than assumed a 
priori. In fact, this is quite analogous to what happens in four-dimensional 
general relativity when one looks for vacuum solutions with only spherical 
symmetry: one ends up with the Schwarschild geometry, which is static.
The above result for empty five-dimensional spacetimes can thus be seen 
as a generalization of Birkhoff's theorem, as it is known in the 
four-dimensional case. 

Since the solutions of the bulk Einstein equations, with the required 
symmetries, are necessarily Sch-AdS, it is easy to infer that the solutions
(\ref{a_bulk}-\ref{n_bulk}), or (\ref{bulk_metric}),
obtained in the particular GN coordinate system correspond to the same 
geometry as  (\ref{adsmetric}) but written in a more complicated 
coordinate system, as can be checked\cite{msm99} by finding the 
 explicit coordinate transformation 
going from (\ref{adsmetric}) to (\ref{metric}).

If the coordinates in (\ref{adsmetric}) are much simpler to describe 
the bulk spacetime, it is not so for the brane itself. 
Indeed, whereas the brane is `at rest' (at $y=0$) in the GN coordinates,
its position $R$ in the new coordinate system,  
will be   in general time-dependent. This means 
that the {\it brane is moving} in the  manifestly static reference frame
(\ref{adsmetric}). 
The trajectory of the brane can be defined 
by its coordinates $T(\tau)$ and $R(\tau)$ given as functions of a 
parameter $\tau$. Choosing $\tau$  to be the proper time 
imposes the condition 
\beq
g_{ab}u^a u^b=- f\dot T^2+{\dot R^2\over f}=-1, 
\label{normalization}
\eeq
where $u^a=(\dot T, \dot R)$ is the brane velocity  and a dot 
stands for  a derivative with respect to the parameter $\tau$.
The normalization condition (\ref{normalization}) yields
$\dot T={\sqrt{f+\dot R^2}/ f}$ and 
the components of the unit normal vector (defined such that $n_au^a=0$ and 
$n_an^a=1$) are, up to a sign ambiguity,
\beq
n_a=\left(\dot R, -{\sqrt{f+\dot R^2}/f}\right).
\eeq
The four-dimensional metric induced in the brane worldsheet is then 
directly given by 
\beq
ds^2=-d\tau^2+R(\tau)^2 d\Omega_k^2,
\eeq
and it is clear that the scale factor of the brane, denoted $a$ previously, 
can be identified with the radial coordinate of the brane $R(\tau)$.

The dynamics of the brane is then obtained by writing the junction conditions
for the brane in the new coordinate system\cite{cr99,kraus99}.
The `orthogonal' components of the extrinsic curvature tensor are given 
by 
\beq
K_{ij}={\sqrt{f+\dot R^2}\over R} g_{ij},
\eeq
which, after insertion in the junction conditions (\ref{junction}),
 implies  for a  mirror symmetric brane (see  Section 9.2 for 
an asymmetric brane)  
\beq
{\sqrt{f+\dot R^2}\over R}={\kappa^2\over 6}\rho_b.
\label{junction_ij}
\eeq
After taking the square of this expression, substituting (\ref{f}) and 
rearranging, we get
\beq
{\dot R^2\over R^2}={\kappa^4\over 36}\rho_b^2+{\Lambda\over 6}
+{\C\over R^4}-{k\over R^2},
\eeq
which, upon the identification between $a$ and $R$, is exactly the 
 Friedmann equation (\ref{fried}) obtained before.  
There is additional information  in the `longitudinal'
 part of the junction conditions. The `longitudinal' component 
of the extrinsic curvature 
tensor is given by 
\beq
K_{\tau\tau}\equiv K_{AB}u^Au^B=u^Au^B\nabla_A n_B=-n_B a^B, 
\eeq
where $a^B$ is the acceleration, i.e. $a^B=u^A\nabla_A u^B$. 
Since $u_B a^B=0$, the acceleration vector is necessarily of the form
$a^B=a n^B$, and $K_{\tau\tau}=-a$. Finally, using the fact that 
${\bf \partial/\partial T}$ is a Killing vector, one can easily show
that 
$a=n_T^{-1}(d u_T/d\tau)$.
Substituting in the junction conditions, one finds
\beq
{1\over \dot R}{d\over d\tau}\left(\sqrt{f+\dot R^2}\right)=
-{\kappa^2\over 6}\left(2\rho_b+3p_b\right)
\label{junction_00}
\eeq
Combining with the first relation (\ref{junction_ij}), it is immediate to 
rewrite the above expression as the traditional cosmological conservation 
equation
\beq
\dot \rho_b +3{\dot R\over R}\left(\rho_b+p_b\right)=0.
\eeq

One has thus established the complete equivalence between the two pictures: 
in the first, the brane sits at a fixed position in a GN coordinate 
system and  while it evolves cosmologically  the metric components evolve with 
time accordingly; in the second, one has a manifestly static bulk spacetime 
 in which the brane is moving, its cosmological 
evolution being simply a consequence of  its displacement in the bulk 
(phenomenom sometimes called `mirage cosmology'\cite{kek99}).
Let us add that the metric (\ref{adsmetric}) describes in principle only one 
side of the brane. In the case of a mirror symmetric brane, the complete 
spacetime is obtained by gluing,along the brane worldsheet,
 two copies of a portion of Sch-AdS spacetime. 
For an asymmetric brane, one can glue two
(compatible) portions from different Sch-AdS spacetimes, as illustrated 
in  Section 9.2.

\section{Inflation in the brane}
The archetypical scenario of nowadays early universe cosmology is 
inflation. Since the infancy of brane cosmology, brane
inflation  has attracted some attention \cite{dt98,kl99,htr00}. 
The simplest way 
to get inflation in the brane is to detune the brane tension from 
its Randall-Sundrum value (\ref{rs}), 
and to take it bigger so that the 
net effective four-dimensional cosmological constant is positive. This
 leads to an exponential expansion in the brane, as illustrated before 
in (\ref{+0}) (we take here $k=0$ and $\C=0$ for simplicity).
In the GN coordinate system, the metric corresponding to this situation 
 \cite{kaloper99} is given by the specialization of the  bulk metric 
(\ref{bulk_metric}) to the case $\eta=\tilde\eta$. The metric 
components $n(t,y)$ and $a(t,y)$ are then separable, i.e. they 
can be written as 
\beq
a(t,y)=a_0(t) \A(y), \quad n=\A(y),
\label{dS}
\eeq
with 
\beq
\A(y)=   \cosh\mu y-\left(1+{\rho\over\sigma}\right) \sinh\mu|y|.
\label{A}
\eeq
Of course, this model is too naive for realistic cosmology since 
brane inflation would never end. Like in standard cosmology, 
one can replace the cosmological constant, 
or here the  deviation of the brane tension from its RS value, 
by a scalar field 
$\phi$ whose potential $V(\phi)$  
behaves, during slow-roll motion, like an effective tension. 
The simplest scenario is to consider a four-dimensional scalar field confined
to the brane\cite{mwbh99}. 
The brane cosmology formulas established above then apply, 
provided one substitutes for the energy density $\rho$ 
and pressure $p$ the appropriate expressions for a scalar field, namely
\beq
\rho_\phi={1\over 2}\dot\phi^2 + V(\phi), 
\qquad p_\phi={1\over 2}\dot\phi^2 - V(\phi).
\eeq
Like in the  analysis for ordinary matter, the   brane inflationary 
scenarios will  be divided into two categories, according to the typical
value of the energy density during inflation:
\begin{itemize}
\item
high energy brane inflation if $\rho_\phi> \sigma$
\item 
low energy brane inflation if $\rho_\phi\ll \sigma$, in which case 
the scenario is exactly similar to four-dimensional inflation.
\end{itemize}
New features appear for the high energy scenarios\cite{mwbh99}:
 for example, it is easier
to get inflation because the Hubble parameter is bigger than the standard 
one, producing a higher friction on the scalar field. 
It enables inflation to take place with potentials usually too steep to 
sustain it \cite{cll00}.
Because of the modified  Friedmann equation, 
the slow-roll conditions are also changed. 

For high energy inflation, the predicted spectra 
for scalar and tensor perturbations are also modified. 
It has been argued \cite{mwbh99}
that, since the scalar field is intrinsically four-dimensional, the 
modification of the scalar spectrum is due simply to the change of the 
background equations of motion.
However, the gravitational wave spectrum requires more care because the 
gravitational waves are five-dimensional objects. This question will be 
treated in detail  in the next section.
Let us also mention another  type of  inflation in brane models, based  
by a bulk five-dimensional scalar field that induces 
inflation within the brane \cite{sasaki}.

One of the main reasons to invoke inflation in ordinary four-dimensional
cosmology  is the 
horizon problem of the standard big bang model, i.e. how to get 
a quasi-homogeneous CMB sky when the Hubble radius size at the time of 
last scattering corresponds to an angular scale of one degree.
Rather than adapt inflation in brane cosmology, one may wonder wether it is
possible to solve the horizon problem altogether without using inflation.
A promising feature of brane cosmology in this respect is the possibility 
for a signal to propagate more rapidly in the bulk than along the brane
\cite{cf99}, thus 
modifying ordinary causality\cite{ishihara}. 
This property 
can also be found in non cosmological brane models, usually   
called Lorentz violation \cite{grojean}.
Unfortunately, in the context of the Sch-AdS brane models presented 
here, a quantitative analysis (for $\C=0$) has shown that 
the difference between the gravitational wave horizon (i.e. for signals 
propagating in the bulk) and the photon horizon (i.e. for signals 
confined to the brane) is too small to be useful as an alternative 
solution to the horizon problem \cite{cl01}.
Note also that the homogeneity problem of standard cosmology, i.e. 
why the universe was so close to homogeneity in its infancy, 
might appear more severe in the braneworld context, where one must explain 
both the homogeneity along the ordinary spatial dimensions and the 
inhomogeneity along the fifth dimension \cite{ck01}.

\section{Cosmological perturbations}
With the  homogeneous scenario presented in section 4 as a starting 
point, one would like to explore the much richer, and much more 
difficult,  question 
of cosmological perturbations and, in particular  investigate whether 
brane cosmology leads to new effects that could be tested in the forthcoming 
cosmological observations, in particular of 
the anisotropies of the Cosmic Microwave Background (CMB).
Brane cosmological perturbations is a difficult subject and although there 
are now many published works\cite{mukohyama,kis00,maartens00,l00a,bdbl00,ks00,bdmp00,l00b,bmw00,ddk00,lmsw00,lmw00,bdw01,grs01,ldcl01,cedric02,rvsd02}
 with various formalisms on this question,  
no observational  signature has yet been  predicted. 
Rather than entering  into the technicalities of the subject, for which 
the reader is invited to consult the original references,  
this section will try to summarize a few  
 results concerning two different, but illustrative, aspects 
of perturbations: on  one hand, the evolution of 
scalar type perturbations {\it on the brane}; on the other hand, the 
production of gravitational waves from quantum fluctuations during an 
inflationary phase in the brane.

\subsection{Scalar cosmological perturbations}
In a metric-based approach, there are  various choices for the gauge 
in which the metric perturbations are defined. We will choose here 
a GN gauge, which has the advantage that the {\it perturbed brane} is still 
positioned at $y=0$ and that only the four-dimensional part of the 
metric is perturbed. One can then immediately identify the value at $y=0$
of the metric perturbations with the usual cosmological metric perturbations
defined by a brane observer~\cite{l00a}.
The most general metric with linear scalar  perturbations about
a FLRW brane is
\begin{equation}
\label{pertmetric}
g_{\mu\nu} = \left[
\begin{array}{ccc}
-(1+2A) & & aB_{|i} \\ && \\
aB_{|j} & & a^2 \left\{ (1+2\R) \gamma_{ij} + 2E_{|ij} \right\}
\end{array}
\right] \,,
\end{equation}
where $a(t)$ is the scale factor and a vertical bar 
denotes the covariant derivative of the three-dimensional metric 
$\gamma_{ij}$.

The perturbed energy-momentum tensor for matter on the brane,
with background energy density $\rho$ and pressure $P$, can be
given as
\begin{equation}
\label{Tmunu}
\tau^\mu_\nu = 
\left[
\begin{array}{ccc}
-(\rho+\delta\rho) & & a(\rho+P) ( v +B )_{|j} \\ &&\\
-a^{-1}(\rho+P)v^{|i} & & (P+\delta P)\delta^i_j + \delta\pi^i_j
\end{array}
\right] \,,
\end{equation}
where
$\delta\pi^i_j=\delta\pi^{|i}{}_{|j}-{1\over3}
\delta^i_j\delta\pi^{|k}{}_{|k}$
is the tracefree anisotropic stress perturbation.
The perturbed quadratic energy-momentum tensor, defined in (\ref{pi}), is
\begin{equation}
\Pi^\mu_\nu =
{\rho\over12}
\left[
\begin{array}{ccc}
-(\rho+2\delta\rho) & & 2a(\rho+P)(v +B )_{|j} \\&&\\
-2a^{-1}(\rho+P)v^{|i} & &
\left\{2P+\rho+2(1+P/\rho)\delta\rho + 2\delta P\right\}\delta^i_j
- (1+3P/\rho)\delta\pi^i_j
\end{array}
\right] \,. 
\end{equation}
The last term on the right hand side of 
 the effective four-dimensional Einstein equations (\ref{einstein4d}) is 
 the projected Weyl tensor $E^\mu_\nu$.  Although it is
due to the effect of bulk metric perturbations not defined on the brane, 
one  can parametrize it as an effective energy-momentum tensor (\ref{Tweyl})
\begin{equation}
T^\mu_{Weyl\ \nu}
 = 
\left[
\begin{array}{ccc}
-(\rho_\E+\delta\rho_\E) & & a\dqE_{|j} \\&&\\
-a^{-1}\dqE^{|i}+a^{-1}(\rho_\E+P_\E)B^{|i} & &
(P_\E+\delta P_\E)\delta^i_j
 + \delta\pi_{\E}{}^{i}{}{}_{j}
\end{array}
\right] \,.
\end{equation}
With these definitions, one can write explicitly the perturbed effective
Einstein equations on the brane, which will look exactly as the 
four-dimensional ones for the geometrical part but with extra terms 
due to $\Pi_{\mu\nu}$ and $T^{Weyl}_{\mu\nu}$ for the matter part.
In addition to the effective four-dimensional Einstein equations, 
the five-dimensional Einstein equations also provide three other equations.
Two of them are equivalent to the conservation of the matter energy
and momentum on the brane, i.e. of the tensor $\tau_{\mu\nu}$. 
The final one yields an equation of
state for the Weyl fluid, which in the 4-dimensional equations
follows from the symmetry properties of the projected Weyl tensor,
requiring $P_\E={1\over3}\rho_\E$ in the background and $\delta
P_\E={1\over3}\delta\rho_\E$ at first order.
The  equations of motion for the effective energy and
momentum of the projected Weyl tensor are provided by the
4-dimensional contracted Bianchi identities, which 
 are {\em intrinsically four-dimensional,} only
being defined on the brane and  not part of the five-dimensional
field equations.
The contracted Bianchi identities ($\nabla_\mu G^\mu_\nu=0$)
and energy-momentum conservation for matter on the brane 
($\nabla_\mu \tau^\mu_\nu=0$) ensure, using Eq.~(\ref{einstein4d}), that
\begin{equation}
\nabla_\mu E^\mu_\nu = \kappa^4\,\nabla_\mu\Pi^\mu_\nu \,.
\end{equation}
In the background, this tells us that $\rho_\E$ behaves like radiation, 
as we knew already,
and for the first-order perturbations we have
\begin{equation}
\label{Econtinuity}
\dot{\delta\rho}_\E + 4H\delta\rho_\E + 4\rho_\E \dot\R
+a^{-1}\left[\nabla^2\delta q_\E
 + {4\over3} \rho_\E \nabla^2 \left(a\dot{E}-{B}\right)\right] = 0
\, .
\eeq
This means that the effective energy of the projected Weyl
tensor is conserved independently of the quadratic energy-momentum
tensor. The only interaction is a momentum transfer~\cite{sms99,maartens00},
as shown by the perturbed momentum conservation equation 
\begin{eqnarray}
\dot{\delta q}_\E &+& 4H\delta q_\E +a^{-1}\left[ {4\over3}\rho_\E A
 + {1\over3} \delta \rho_\E + {2\over 3}
(\nabla^2+3k)\delta\pi_\E \right]=
\cr 
&&{\rho+P \over a\sigma} \left[\delta\rho - 3Ha\delta q
  - (\nabla^2+3k) \delta\pi \right] \,, 
\end{eqnarray}
where  the right hand side represents the momentum transfer
from the quadratic energy-momentum tensor. 
 
It is also possible to construct \cite{lmsw00} gauge-invariant variables 
corresponding to the curvature perturbation on hypersurfaces of
uniform density, both for the brane matter energy density 
and for the total effective energy density (including the quadratic terms 
and the Weyl component). These quantities are extremely useful because 
their evolution on scales larger than the Hubble radius can be solved easily.
However, their connection to the large-angle 
CMB anisotropies involves the knowledge of anisotropic stresses due to the 
bulk metric perturbations\cite{lmsw00}. 
This means that  {\it for a quantitative prediction 
of  the CMB anisotropies, even at large scales,  one needs to determine 
the evolution of the bulk perturbations}.

In summary, we have obtained a set of equations for the brane linear 
perturbations,  
where one recognizes the ordinary cosmological  equations 
but modified by   two  types of corrections:
\begin{itemize}
\item modification of the  homogeneous background coefficients due to the 
additional $\rho^2$ terms in the Friedmann equation.   These  corrections are  
negligible in the low energy  regime $\rho\ll\sigma$.
\item presence of source terms in the  equations. 
These terms come from the bulk perturbations and cannot be determined solely 
from the evolution inside the brane. To determine them, one must solve 
the full problem in the bulk (which also means to specify some initial 
conditions in the bulk). In the effective four-dimensial perturbation
equations, these 
terms from the fifth dimension appear like external source terms, in a way 
somewhat similar to  
the case  of ``active seeds'' due to  topological defects. 
\end{itemize}

\subsection{Inflationary production of gravitational waves}
Let us turn now to another facet of the brane cosmological perturbations:
their production during a brane inflationary phase\cite{lmw00}. 
We concentrate on the tensor perturbations, which are subtler 
than the scalar perturbations, because gravitational waves 
have   an extension in the fifth dimension. 
The brane gravitational waves can be defined by a perturbed 
 metric of the form 
\beq
ds^2=-n^2 dt^2
+a^2\left[ \d_{ij}+E_{ij}^{TT}\right]dx^i dx^j
+dy^2,
\eeq
where the  `TT' stands for transverse traceless. The linearized 
Einstein equations for the metric perturbations give a wave equation, 
which reads in Fourier space  
($E_{ij}=E \ e^{i \vec k. \vec x}\  e_{ij}$) 
\beq
\ddot{E}+3H_0\dot{E}+{{\vec k}^2\over a_0^2}E=\A^2 E''+4\A \A'E'\,.
\label{E}
\eeq
where $\A$ is defined in (\ref{A}).
This equation being separable, one looks for solutions 
 $E=\varphi_m(t) \H_m(y)$, where the time-dependent part obeys
 an ordinary FLRW Klein-Gordon equation
while  the $y$-dependent part must satisfy 
the Schr\"odinger type equation 
\beq
\label{SE}
 {d^2\Psi_m\over dz^2} - V(z)\Psi_m =-m^2 \Psi_m \,,
\end{equation}
after introducing   the new variable  
$z-z_b=\int_0^y d\tilde y/\A(\tilde y)$ 
(with $z_b=H_0^{-1}\sinh^{-1}(H_0/\mu)$)
and the new function
 $\Psi_m= \A^{3/2}\H_m$.
The potential is given by 
\beq
V(z)= {15H_0^2 \over 4\sinh^2(H_0 z)} +
{\textstyle{9\over4}}H_0^2
- 3\mu\left[1+{\rho\over\sigma}\right] \delta(z-z_{\rm b}) \,.
\eeq
\begin{figure}
\epsfbox{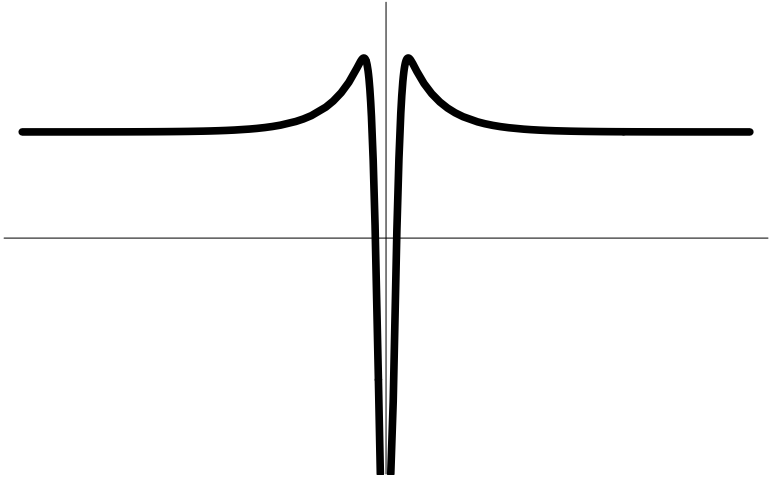}
\caption{Potential for the graviton modes in a de Sitter brane}
\end{figure}
The non-zero value of the Hubble parameter signals the presence 
of a gap\cite{gs99}
between the zero mode ($m=0$) and the continuum of Kaluza-Klein modes ($m>0$). 
The zero mode corresponds to $\H_0= const$ and the constant is determined 
by  the normalization $2 \int_{z_{\rm b}}^\infty |\Psi_0^2| dz=1$. 
One finds  
\beq
\H_0=  \sqrt{\mu}\,\,F\!\left({H_0/\mu}\right)\,, \qquad
F\!\left(x\right) =\left\{ \sqrt{1+x^2} - x^2 \ln \left[ {1\over
x}+\sqrt{1+{1\over x^2}} \right] \right\}^{\!\!-1/2}
\!.
\eeq
Asymptotically, $F\simeq 1$ at low energies, i.e. for $H_0\ll \mu$, 
and  $F\simeq \sqrt{3H_0/(2\mu)}$ at very high energies, i.e. for $H_0\gg \mu$.
One can then evaluate the vacuum quantum fluctuations of the zero mode 
by using the standard  canonical quantization. To do this 
explicitly, one writes the five-dimensional action for gravity at 
second order in the perturbations. Keeping only the zero mode and integrating
over the fifth dimension, one obtains  
\beq
S_{\rm g}= {1\over8\kappa^2}
\, \sum_{+,\times}\, \int d\eta\, d^3\vec{k}\,a_0^2\left[
\left({d\varphi_0 \over
d\eta}\right)^2+k^2{\varphi_o}^2\right] \,, 
\eeq
This has the standard form of a massless graviton in
four-dimensional cosmology, apart from
the overall factor $1/8\kappa^2$ instead of $1/8\kappa_4^2$. It
follows that quantum fluctuations in each 
polarization, $\varphi_0$,
have an amplitude of $2\kappa (H_0/2\pi)$ on super-horizon scales. Quantum
fluctuations on the brane at $y=0$, where $E_0=\H_0\varphi_0$, thus have the
typical amplitude\cite{lmw00}
\beq
{1\over 2\kappa_4}\delta E_{\rm brane}=\left({H_0\over 2\pi}\right) F(H_0/\mu)
\eeq
The same result can be obtained in a bulk-based approach \cite{grs01}.

At low energies, $F=1$ and 
one  recovers exactly the usual four-dimensional result 
but at higher energies the multiplicative factor $F$ provides an 
{\it enhancement of the gravitational wave spectrum amplitude 
with respect to the four-dimensional result}. However, comparing  
this with the amplitude for the scalar spectrum\cite{mwbh99},
one finds that, at high energies ($\rho\gg\sigma$), the {\it 
tensor over scalar 
ratio is in fact suppressed with respect to the four-dimensional ratio}.
An open question is how the gravitational waves will evolve during the 
subsequent cosmological phases, the radiation and matter eras.

\section{Non-empty bulk}
Out of simplicity, a majority of  works in brane cosmology have focused 
on  a brane-universe
embedded in a five-dimensional {\it empty} bulk, i.e. with only gravity 
propagating in the bulk.  
But many works have also considered extra fields in the bulk, 
either motivated by M/string theory \cite{low,reall98,lidsey01}, or simply in a purely 
phenomenomogical approach \cite{mw00,mb00}, in some cases with the objective 
to solve some standing problems such as the cosmological constant problem 
\cite{selftuning} or more specific problems like the question of the radion 
stabilization\cite{gw99,stabilization}.
The most common extra field considered in the literature is not 
surprisingly the scalar field although one can find other generalizations 
such as including gauge fields \cite{carter,gqtz01}.
We will consider here only models   with  a bulk scalar field and start  
from the  action
\beq
{\cal S} = \int d^5 x \,\sqrt{-g}\,\left[ \frac{1}{2\,
\kappa^2}\,{}^{(5)}R - \frac{1}{2} \,(\nabla_A \phi)\nabla^A \phi - V(\phi)\,\right] 
+ \int_{brane} d^4 x \, {L}_{m}[\varphi_m,\tilde h_{\mu\nu}],
\label{action}
\eeq
where it is assumed that the four-dimensional metric $\tilde h_{\mu\nu}$
which is minimally 
coupled to the four-dimensional matter fields $\varphi_m$ in the 
brane, is conformally related to the induced metric $h_{\mu\nu}$, i.e.
\beq
\tilde h_{\mu\nu}=e^{2\xi(\phi)}h_{\mu\nu}.
\eeq
In the terminology  of scalar-tensor theories, 
one would 
say that $h_{\mu\nu}$ corresponds to the Einstein frame while 
$\tilde h_{\mu\nu}$ corresponds to the Jordan frame. One can also  define 
two brane matter energy-momentum  tensors, $T_{\mu\nu}$ and 
$\tilde T_{\mu\nu}$, respectively with respect to $h_{\mu\nu}$ and
$\tilde h_{\mu\nu}$, 
\beq
T^{\mu\nu}={2\over\sqrt{-h}}{\delta L_m\over \delta h_{\mu\nu}},
\quad
\tilde T^{\mu\nu}={2\over\sqrt{-\tilde h}}
{\delta L_m\over \delta \tilde h_{\mu\nu}},
\eeq
and they  are related by $\tilde T^\mu_\nu=e^{-4\xi} T^\mu_\nu$. 
The corresponding
energy densities, $\rho$ and $\tilde \rho$, and pressures, $p$ and $\tilde p$, 
will be related by the same factor $e^{-4\xi}$.
Variation of the action (\ref{action}) with respect to the metric $g_{AB}$ 
yields 
the five-dimensional Einstein equations (\ref{einstein}), 
where, in addition to the 
(distributional) brane energy-momentum, there is now the scalar field
energy-momentum tensor 
\beq
T^\phi_{AB}=\partial_A\phi\partial_B\phi-g_{AB}\left[{1\over 2}(\nabla_C\phi)
(\nabla^C\phi)+V(\phi)\right]. \label{mom-tensor}
\eeq
Variation of (\ref{action}) with respect to $\phi$ 
yields the equation 
of motion for the scalar field, 
which is of the Klein-Gordon type, with  
a distributional source term since  the scalar field is coupled to the brane
via $\tilde h_{\mu\nu}$.
This implies that there is an additional junction condition, now involving  
 the scalar field at the brane location and  which is of the form
\beq
\left[n^A\partial_A\phi\right]=-\xi'e^{4\xi}\tilde T=-\xi' T,
\label{junction_phi}
\eeq
where $T=-\rho+3P$ is the trace of the energy-momentum tensor. 
In summary, we now have a much more complicated system than for the 
empty bulk, since, in addition 
to the brane dynamics, one must solve for the dynamics 
of the scalar field. Although a full treatment would require a numerical 
analysis, 
a  few interesting analytical  solutions have been derived in the literature
\cite{bd,fkv01,lr01,davis01,charmousis01,ftw01}, 
out of which we give below two examples.

\subsection{Moving brane in a static bulk}
Taking example on the empty bulk case, where the cosmology in the brane
is induced by the motion of the brane in a static bulk, 
it seems natural to look for  brane cosmological solutions 
corresponding to  a moving brane in a {\it static} bulk geometry with 
a  static scalar field.
In the  case of  an exponential 
potential, 
\beq
V(\phi)=V_0\exp\left(-{2\over \sqrt{3}}
\lambda \kappa\phi\right),
\eeq
it turns out that there exists a simple class of  static solutions\cite{static}
(the full set of static solutions is larger\cite{charmousis01}),
which correspond to a metric of the form  
\beq
ds^2=-h(R)dT^2+{R^{2\lambda^2}\over h(R)}dR^2+R^2 d{\vec x}^2,
\label{dil_stat}
\eeq
with 
\beq
h(R)=-{\kappa^2 V_0 /6\over 1-(\lambda^2/4)}R^2-\C R^{\lambda^2-2},
\eeq
where $\C$ is an arbitrary constant, 
and a scalar field given by
\beq
{\kappa\over \sqrt{3}}\phi=\lambda\ln(R).
\eeq
The next step is then to insert  a moving (mirror-symmetric) brane 
in the above bulk configuration\cite{cr99,lr01}. This is possible if  
 the three junction conditions, two  for the 
metric and one for the scalar field, are satisfied. The two metric 
junction conditions are easily derived by repeating the procedure given 
in section 5    for the metric (\ref{dil_stat}) and will be generalizations
of  (\ref{junction_ij}) and (\ref{junction_00}). The condition on the 
scalar field comes from (\ref{junction_phi}).

These three junction conditions can be rearranged to give the 
following three relations:
\begin{itemize}  
\item  a generalized  Friedmann equation,
\beq
H^2={\kappa^2\over 36}\rho^2- {h(R)\over R^{2+2\lambda^2}}=
{\kappa^2\over 36}\rho^2+{\kappa^2 V_0 /6\over 1-(\lambda^2/4)}R^{-2\lambda^2}
+\C R^{-4-\lambda^2},
\eeq
\item a (non-) conservation equation for the energy density,
\beq
\dot\rho+3H(\rho+p)=(1-3w)\xi' \rho \dot \phi,
\eeq
\item 
a constraint on  the brane matter equation of state, which must be related 
to the conformal coupling according to the expression 
\beq
3w-1={\kappa\over\sqrt{3}}{\lambda\over \xi'}.
\eeq
\end{itemize}
The latter constraint  can  accomodate a constant $w$, provided
the conformal coupling is of the form  $\xi(\phi)=\xi_1\phi$, but it is not 
very sensible for realistic cosmological scenarios.
If there is no coupling between the brane and the scalar field, i.e.
$\xi'=0$, then  $\lambda$ must vanish and one 
recovers the familiar brane solutions with an empty bulk where $w$ is 
unconstrained.
It is also worth noticing that, when 
 going to the Jordan frame, i.e. using the 
scale factor $\tilde a=e^\xi a$ and the time $\tilde t$, such that 
$d\tilde t= e^\xi dt$, the conservation equation  for the energy 
density transforms  into its familiar form
\beq
{d\tilde\rho\over d\tilde t}+3\tilde H\left(\tilde\rho+\tilde p\right)=0.
\eeq

\subsection{Non-static bulk solutions}
In contrast with the empty bulk case, where the required symmetries impose
the bulk to be static, there now exist  cosmological solutions with non static 
bulk configurations because  the generalized Birkhoff's theorem 
no longer applies. 
The example below provides an explicit illustration of this fact.
Still for an exponential potential 
\beq
V=V_0\exp\left(\alpha {\kappa\over\sqrt{3}}\phi\right),
\eeq
a solution of the Einstein/Klein-Gordon bulk equations is given by\cite{lr01}
the metric 
\beq
ds^2={18\over \kappa^2 V_0}e^{\alpha^2T} \, e^{-\alpha\sqrt{\alpha^2- 4} R}
\left(-dT^2+dR^2\right) + e^{4T} d{\vec x}^2,
\eeq
and the scalar field configuration 
\beq
\phi=\alpha^2T -\alpha\sqrt{\alpha^2- 4} R.
\eeq
It is possible to embed a (mirror) symmetric brane in this spacetime, with 
an equation of state $p_b=w\rho_b$ ($w$ constant). This  leads, 
 for a brane observer in the Einstein frame, to a cosmology with  
the  power-law expansion
\beq
a(t)\sim t^p, \qquad p={1\over 3(1+w)}\left[1+{4(2+3w)\over\alpha^2}\right],
\eeq
with  $p$  between the values $p=1$ and $p=1/(3(1+w))$, 
since the solution is defined only for $\alpha^2\geq 4$; 
  the scalar field in the brane is given by 
\beq
{\kappa\over\sqrt{3}}\phi_0=
-{6\alpha(2+3w)\over \alpha^2+4(2+3w)}\ln a 
\eeq
and the energy density is proportional to Hubble parameter,
\beq
{\kappa^2\over 6}\rho_b={\alpha\sqrt{\alpha^2- 4}\over 2\left(\alpha^2
+4(2+3w)\right)}H.
\eeq
It is then straightforward to go to the Jordan frame. As in the previous 
example, these solutions are very contrived and, moreover, they can be 
of interest only for the very early universe because of their Brans-Dicke 
nature.

\section{Multi-brane cosmology}
The last part of this review  will be devoted to 
scenarios where several branes are involved. So far, we have concentrated 
our attention on a single brane, supposed to correspond to our accessible 
universe. Nothing forbids however the presence of other branes which, 
because of the required cosmological symmetries, would be `parallel' or 
`concentric' with respect to our brane-universe.  
A very important property in the case of an empty bulk spacetime 
(apart the branes themselves of course), which is not always well appreciated,
is that {\it the cosmology in one brane is completely independent of 
the evolution of any other brane}.  The only influence of another brane
on, say, our brane-universe is contained in the {\it single constant} $\C$, 
which appears in the Friedmann equation (\ref{fried}). But $\C$ is just 
determined {\it at any given time} by a kind of compatibility condition 
between any two branes and the empty spacetime separating them. This 
compatibility condition is then automatically satisfied at subsequent times 
whatever the evolution of each brane.
Below, we first study the cosmology of two parallel (mirror-symmetric)
branes.  We then consider the  general problem of collisions in 
a multi-brane system.
  
\subsection{Two-brane system}

\def\ro{{\rho_1}}
\def\y{{\cal R}}
\def\dy{{\delta\y}}

Sometimes motivated 
by the old prejudice that extra dimensions must be compact 
(although this is no longer necessary as illustrated by the 
Randall-Sundrum model), a lot of attention has been paid to two-brane
systems. This is in particular the case with the first Randall-Sundrum 
model, which was supposed to solve the hierarchy problem, although 
this model is incompatible with our gravity\cite{gt99}
 and should be completed for 
instance with a stabilization mechanism \cite{gw99} in order to be 
phenomenologically valid\cite{tm00}. This subsection will present a 
few results concerning systems with two parallel (planar) branes 
separated by vacuum (with a negative cosmological constant if one 
wishes to recover standard cosmology in our brane). The two branes 
will be assumed to be mirror symmetric.

We will adopt the brane point of view and use a GN coordinate system 
based on our brane-universe. As already shown, the metric then reads
\beq
  ds^2=
-n(t,y)^2 dt^2+a(t,y)^2 d\Sigma^2+ dy^2,\label{metric2}
\eeq
where the explicit expressions for $a(t,y)$ and $n(t,y)$ are given in 
(\ref{a_bulk}) and (\ref{n_bulk})). While 
$y=0$ is the position of our brane-universe,
the position of the second brane, at any time $t$, will be expressed 
in terms of its coordinate $y=\y(t)$.  It represents the relative distance 
between the two branes, as measured by an observer at rest with respect to 
our brane-universe, and will be called the {\it cosmological radion}. 
Using the definition of the extrinsic curvature tensor (\ref{extrinsic}), 
one can  compute the junction conditions for the second brane\cite{bdl01}. 
The `orthogonal' components 
give 
\beq
\frac{a^\prime}{a}+ \frac{\dot a}{a}\frac{\dot \y}{n^2} = 
 {\kappa^2\over 6}
\ro\left(1-\frac{\dot {\y}^2}{n^2}\right)^{1/2},
\label{junction_a_R}
\eeq
 whereas
the `longitudinal' component yields
\beq
\frac{\ddot \y}{n^2} + \frac{n^\prime}{n}
\left(1- 2 \frac{\dot{\y}^2}{n^2}\right)  = -{\kappa^2\over 6} 
\left(2\ro + 3P_1)
\right)\left(1-\frac{\dot \y^2}{n^2} \right)^{3/2}.
\label{junction_n_R}
\eeq 
The quantities $\rho_1$ and $P_1$ refer to the {\it total} energy 
density and pressure of the second brane.
All the $y$-dependent quantities in the above equations are of course
evaluated at $y=\y(t)$. 

If the second brane is at rest, i.e. $\dot\y=0$, then 
the junction conditions are, with the important exception of the sign, 
the same relations as written before in (\ref{junction_a_n}). 
The equilibrium condition thus consists 
of two relations relating the energy densities and pressures in the two 
branes, involving  as well 
 the value of the radion $\y_{eq}$. They  can be rewritten 
in the very simple form
\beq
\theta_0+\theta_1=\tilde\theta_0+\tilde\theta_1=\mu\y_{eq},
\eeq
where $\eta_i=\tanh\theta_i$ and ${\tilde\eta_i}=\tanh{\tilde\theta_i}$
($\eta$ and $\tilde\eta$ have been defined in (\ref{eta})).
Note that in the Randall-Sundrum two-brane configuration, 
where the two branes are at rest 
relative to each other, $\theta_0=\tilde\theta_0=+\infty$ and 
$\theta_1=\tilde\theta_1=-\infty$, and $\y_{eq}$ can take any value.

If the second brane is not at rest, then the above relations  
(\ref{junction_a_R}) and (\ref{junction_n_R})
represent generalized junction conditions when the brane moves with 
respect to the coordinate frame. It is also instructive to consider 
linearized radion fluctuations about an equilibrium value, i.e. 
$\delta\y=\y-\y_{eq}$. Combining the two linearized junction conditions,
one can obtain 
\beq 
\frac{\ddot\dy}{n^2} +3\frac{\dot a}{a}\frac{\dot \dy}{n^2}
+m_{eff}^2\dy=
-{\kappa^2\over 6} \delta T_1, 
\label{tracefluct}
\eeq
where the effective square mass is given by
\beq
m^2_{eff}=\mu^2\left(4-3\eta_1^2-\tilde\eta_1^2\right).
\eeq
This equation for the radion allows one to make the connection with the 
traditional view that the radion should be seen, from the four-dimensional 
point of view, as a scalar field coupled to the trace of energy-momentum
tensor. Note that a  RS brane corresponds to a vanishing radion mass , whereas 
 the radion is unstable for a de Sitter brane and 
stable for an AdS brane \cite{radion}.

It is sometimes  useful to derive from a model 
with extra-dimensions an effective four-dimensional theory by integrating 
out the extra degrees of freedom. The  braneworld models are however 
very particular, in contrast with the traditional Kaluza-Klein type models, 
because matter is confined on branes. Let us consider a cosmological  
 two-brane system with two mirror symmetric branes.
  As stressed earlier, an important property 
at the homogeneous level is that 
the second brane  {\it does not influence} our brane-universe (and 
reciprocally) with  the exception of the constant $\C$, which can 
be reinterpreted in terms of the second brane, but which does not 
change with time. 
This non-influence can seem at odds with the usual intuition built 
on an effective four-dimensional description, where the Friedmann equation
is expected to involve the energy densities of both branes. We will give below
a brief idea of    how the two viewpoints can be reconciled.

In order to construct a four-dimensional effective action 
describing the two-brane system, one can  start from the full 
five-dimensional action, assuming for simplicity $k=0$ and $\C=0$, and 
that the matter in 
each brane is a pure tension, in which case both branes undergo inflation\cite{kk99}.
 One  substitutes the $y$-dependence of the 
bulk fields given in this case (with $k=0$ and $\C=0$ for simplicity)
by  (\ref{dS}-\ref{A}) and $n=N(t)\A(y)$. The   integration over 
 the fifth dimension yields  the following 
(homogeneous) four-dimensional action 
\cite{ls01}
 \begin{eqnarray}
S=\frac{1}{\kappa^2\,\mu}\int d^4x\,Na_0^3\,&& \left[{}^{(4)}R\,\psi_1(\R)
+12\,\mu^2\,\psi_2(\R)+6\,\mu\,\A_1^2\,\frac{\dot a_0}{N\, a_0}\,\frac{\dot{\R}}{N} +\right. \cr
&&\left.+3\,\mu\,\A_1^3\,\left(\frac{\dot a_0}{N\, a_0}+\mu\,{\A_1'\over \A_1}\frac{\dot\R}{N}\right)\,
\ln\left({N\,\A_1-\dot\R\over N\,\A_1+\dot\R}\right)+\right.\cr 
&&\left.-\kappa^2\,\mu\,\sigma_0
- \kappa^2\,\mu\,\sigma_1\,\A_1^4\sqrt{1-\frac{\dot\R^2}{N^2\,\A_1^2}}\, 
\right]\,\,,
\label{action_4d}
\end{eqnarray}
with $\A_1\equiv \A(\mu\R)$~and $\A'_1\equiv \A'(\mu\R)$, while the $\psi$'s
are dimensionless  functions of $\mu\R$ defined by
\beq
\psi_1(\R)=\int_0^{\mu\R} d\xi\, \A\left(\xi\right)^2, \,\,\,
\psi_2(\R)=\int_0^{\mu\R} d\xi\, \A\left(\xi\right)^2\,
\left(\A'\left(\xi\right)^2+ \A\left(\xi\right)^2\right)\,\,,
\label{psi}
\eeq
and ${}^{(4)}R$ is the (homogeneous) four--dimensional Ricci scalar
\beq
{}^{(4)}R=\frac{6}{N^2}\,\left(\frac{\dot{a}_0^2}{a_0^2}+\frac{\ddot{a}_0}{a_0}-
\frac{\dot{a}_0}{a_0}\,\frac{\dot{N}}{N}\right)\,\,.
\eeq
The variation with respect to $N$ of this action, yields a Friedmann-like 
equation which reads
\beq
H_0^2\,\psi_1+2\mu^2\psi_2+
\mu\left(H_0+\mu\,{\A_1'\over \A_1}\,\dot \R\right){\A_1^2\,\dot\R\over 1-(\dot\R/\A_1)^2}
={\kappa^2\mu\over 6}\,\sigma_0+{\kappa^2\mu\over 6}\,\sigma_1{\A_1^4\over 
\sqrt{1-(\dot\R/\A_1)^2}}, \label{friedmann}
\eeq
setting $N=1$ after variation.
What is worth noticing is that this equation involves the energy densities 
in a linear way, and not in a quadratic way as is characteristic of 
brane cosmology. The solution to this apparent paradox 
lies in the fact that  the equation also involves the radion.  As seen 
above, the junction condition (\ref{junction_a_R}) for 
 the second brane provides a relation between the energy 
densities of the two branes and the radion. It turns out that,  
using this junction condition, one can simultaneously get rid of the 
radion and of the energy density $\sigma_1$ and recover the `usual'
unconventional Friedmann equation quadratic in $\sigma_0$. But, if one ignores
the junction condition, one will in general lose some information about 
the fifth dimensional aspect of the problem and find solutions which are 
not physical. This is only in the low energy limit, i.e. 
near the RS configuration, where the effective potential for the radion is 
quasi-flat, and for small radion velocities, that the four-dimensional 
action yields the correct solutions. One must be therefore very cautious
when trying to infer the cosmological behaviour of an intrinsically 
five-dimensional model from its effective four-dimensional description.

\subsection{Collision of branes}
As soon as the spacetime contains several branes and that these 
branes move with respect to each other, they might {\it collide}.
 A fascinating possibility, which has been actively explored  
recently\cite{kost,bucher01,git01}, 
is that the Big-Bang is such a brane collision.
 Rather than  entering into the details of these various models, 
let us point out here  
a simple and general analysis   \cite{lmw01} of the 
collision of (parallel or concentric) branes separated by  
vacuum, i.e. 
branes separated   by patches of Sch-AdS spacetimes (allowing for different 
Schwarschild-type mass and cosmological constant in each region) with the 
metric (\ref{adsmetric}). Although we are interested here by $3$-branes 
embedded in a five-dimensional spacetime,  this analysis is immediately 
applicable to the case of $n$-branes moving in a $(n+2)$-dimensional 
spacetime, with the analogous symmetries.

To analyze the collision, it is
 convenient to introduce an angle $\alpha$, which characterizes 
the motion of the brane with respect to the coordinate system 
(\ref{adsmetric}),  defined 
by 
\beq
\label{alpha}
\alpha=\sinh^{-1}(\epsilon\dot R/\sqrt{f}),
\eeq
where $\epsilon=+1$ if $R$ decreases from ``left'' to ``right'',  $\epsilon=-1$
otherwise.
Considering   a collision involving a total number of $N$ branes, 
both ingoing and outgoing, thus separated by $N$ spacetime regions 
\begin{figure}
\epsfbox{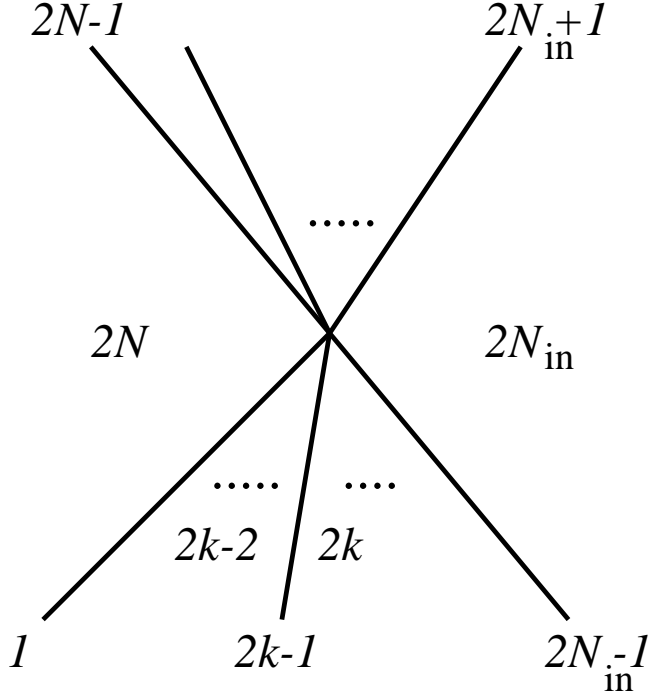}
\caption{Collision of $N_{\rm in}$ ingoing
  branes, yielding $N_{\rm out}=N-N_{\rm in}$ outgoing branes. Odd
  integers denote branes and even integers denote regions in between.}
\end{figure}
one can  label alternately branes
and regions by integers, starting from the leftmost ingoing brane and
going anticlockwise around the point of collision (see Figure).
 The branes will
thus be denoted by odd integers, $2k-1$ ($1\le k \le N$), and the
regions by even integers, $2k$ ($1\le k \le N$).  Let us introduce, as
before, the angle $\alpha_{2k-1|2k}$ which characterizes
the motion of
the brane $\B_{2k-1}$ with respect to the region $\R_{2k}$,
and which is defined by
\beq
\sinh\alpha_{2k-1|2k}={\epsilon_{2k}\dot R_{2k-1}\over \sqrt{f_{2k}}}.
\label{alpha_k}
\eeq
Conversely,  the motion of the region $\R_{2k}$
with respect to the brane by the Lorentz angle
$\alpha_{2k|2k-1}=-\alpha_{2k-1|2k}$.
It can be shown that the junction conditions for the branes can be written 
in the  form
\beq
\label{junction2} \tilde\rho_{2k-1} \equiv \pm{\kappa^2\over 3}\rho_{2k-1} R=
\epsilon_{2k}\sqrt{f_{2k}}\exp{(\pm\alpha_{2k-1|2k})} 
 \ - \epsilon_{2k-2}\sqrt{f_{2k-2}}
\exp{(\mp\alpha_{2k-2|2k-1})} \label{rho},
\eeq
with the plus sign for ingoing branes ($1\le k\le N_{in}$), the minus 
sign for outgoing branes ($N_{in}+1\le k \le N$).
 An outgoing positive energy density brane
thus has the same sign as an ingoing negative energy density
brane.

The advantage of this formalism becomes obvious when one writes 
the {\it geometrical   consistency relation} that expresses the matching 
of  all branes and spacetime regions around the collision point. 
In terms of the angles defined above, it reads simply
\beq
\sum_{i=1}^{2N} \alpha_{i|i+1}=0. \label{collision}
\eeq
Moreover, introducing the generalized angles 
$\alpha_{j|j'}=\sum_{i=j}^{j'-1}\alpha_{i|i+1}$, 
if $j<j'$, and $\alpha_{j'|j}=-\alpha_{j|j'}$,
the sum rule for angles (\ref{collision}) combined with the junction 
conditions (\ref{junction2}) leads to the laws of  energy conservation 
and momentum conservation. 
The energy conservation law reads
\beq
\sum_{k=1}^N\tilde\rho_{2k-1}\gamma_{j|2k-1}=0,
\eeq
where $\gamma_{j|j'}\equiv \cosh\alpha_{j|j'}$ corresponds to the Lorentz
factor between the brane/region $j$ and the brane/region $j'$ and can be
obtained, if $j$ and $j'$ are not adjacent, by combining all intermediary
Lorentz factors (this is simply using the velocity addition rule of
special relativity).
The index $j$ corresponds to the reference frame with respect to which
the conservation rule is written.
Similarly, the {\it momentum conservation law} in the $j$-th reference frame
can be expressed in the form
\beq
\sum _{k=1}^N\tilde\rho_{2k-1}\gamma_{2k-1|j}\beta_{2k-1|j}=0,
\eeq
with $\gamma_{j|j'}\beta_{j|j'}\equiv \sinh\alpha_{j|j'}$.
One thus obtains, just from geometrical considerations, conservation laws
 relating  the brane energies densities and velocities before and after 
the collision point.
These results apply to any collision of branes in vacuum, with the appropriate 
symmetries of homogeneity and isotropy. An interesting development would be 
to extend the analysis to branes with small perturbations and investigate 
whether one can find scenarios which can produce quasi-scale invariant 
adiabatic spectra, as seems required by current observations.

\section*{Acknowledgements}
I would like to thank the organizers of the brane cosmology workshop
at the Yukawa Institute for inviting me to a very  stimulating meeting 
 (and also for their financial support). 
Let me also thank    Genevi\`eve and Thierry Pichevin for 
their warm hospitality in Pospoder, where the first part of this 
review was written. I am also grateful to my new-born son Nathan for his
approving  silence during the last stage of this work.

\end{document}